# The effect of motion of respiratory droplets on propagation of viral particles

## Z. Ugulava[1], Z. N. Osmanov[1,2]*

[1,] School of Physics, Free University of Tbilisi, 0183, Tbilisi, Georgia
[2] E. Kharadze Georgian National Astrophysical Observatory, Abastumani 0301, Georgia

e-mail: zugul20@freeuni.edu.ge, z.osmanov@freeuni.edu.ge

*Corresponding Author: z.osmanov@freeuni.edu.ge

*Abstract*— In this work we investigate viral load propagation due to liquid droplets expelled during respiratory actions. We describe a mechanism of the transmission of such evaporating system and analyze dependence on several ambient parameters for different respiratory processes, such as sneezing, coughing and speaking. Our study has found that a significant amount of viral load and droplets from the respiratory fluid might transfer beyond the customary $2m$ distance, understanding of which might give us a better insight about regulations and risks posed by an infected individual.

## I. INTRODUCTION

As the world is immersed in global pandemic, several studies have been carried out to fully understand the mechanism of propagation and translation of infectious particles. Along with any respiration we exhale a large number of viral particles, therefore analyzing physics behind it could help us issue improved social regulations to minimize an effect of infection. Understanding and describing respiratory actions can be challenging. There have been different types of approaches towards the problem. Some of them consider the gravity induced free fall of the droplet, with a constant Stokes drag coefficient, estimating and comparing total airborne time of a falling droplet to its evaporation time-scale [1–3]. Others have attempted to adapt various types of respiration into a moving jet, solving the motion characteristics via analyzing the flow of such bodies [4–8], however as the liquid starts evaporating in atmosphere, droplets shrink and therefore, the drag coefficient will correspondingly reduce, which might significantly change the overall dynamics. Apart from that, even initially the Reynolds numbers of droplets are so high that the Stokes' formula is not valid any more.

In this paper we study the motion of the liquid droplets, taking into account the effect of evaporation and the varying drag coefficient, which when applied to a set of generated respiratory droplets gives us an insight about a distance traveled and transferred viral load in real-life scenarios for different kinds of face-to-face respiration forms such as sneezing, coughing and speaking loudly.

The paper is organized in the following way. In section 2 we present a theoretical model of the dynamics of droplets, in section 3 we explain modelling methods and finally in section 4 we discuss our results and summarize them.

## II. MAIN CONSIDERATION

In this section we study how the droplets move in a typical ambient, by means of the gravity force, buoyancy and the drag force, by taking into account the process of evaporation of the droplet.

### 2.1. Equations of motion

If we estimate a mean distance $\langle l \rangle$ between average sized droplets with mean radius $\langle r \rangle = 30\mu m$, at concentration of $n_c = 6.25 \cdot 10^8 \, m^{-3}$, we obtain a significantly larger length-scale $\langle l \rangle = n_c^{-1/3} = 1.16 \cdot 10^{-3} \, m$ compared to the actual droplet radius, indicating that droplets do not intervene with each other during the motion. We consider a droplet exiting from the mouth orifice, with a certain initial velocity vector $\vec{\upsilon}_0$. Gravity and buoyant forces are aligned vertically, whereas the drag force due to air resistance is always directed in a direction opposite to the velocity vector. Then, the vector equation governing the dynamics of droplets is given by

$$\frac{d\vec{P}}{dt} = m\vec{g} + \vec{F}_b + \vec{F}_d \qquad (1)$$

where $\vec{P}$ and $m$ are the droplet's momentum and mass respectively, $\vec{g}$ is the free fall acceleration,

$$\vec{F}_b = -\rho_0 V \vec{g} \qquad (2)$$

represents buoyancy,

$$\vec{F}_d = -\frac{1}{2} C_d \pi R^2 \upsilon \vec{\upsilon} \quad (3)$$

is the drag force, $\rho_0$ is the density of the ambient and $V$ and $R$ denote respectively the volume and radius of the droplet. Here $C_d$ is the drag coefficient which unlike the models considered in [1–3] depends on the Reynolds number $\mathrm{Re} = 2\upsilon R \rho_0 / \mu_0$ and viscosity ratio $(\lambda \equiv \mu / \mu_0)$ of media inside and outside of the liquid droplet because for the typical velocities of droplets the flow is turbulent, correspondingly the Reynolds number of such droplets, expelled mostly vary within the range [200,1000] and ratio $\lambda \gg 2$, so we adopted Feng and Michaelides' [9] drag model, which follows:

$$C_d = \frac{68}{\lambda+2}\mathrm{Re}^{-2/3} + \frac{\lambda-2}{\lambda+2}\frac{24}{\mathrm{Re}}\left(1+\frac{1}{6}\mathrm{Re}^{1/3}\right) \quad (4)$$

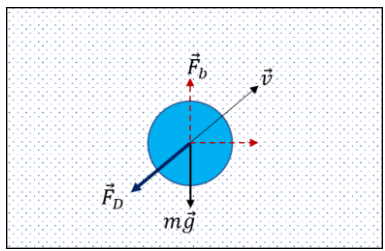

Figure 1: Visual representation of the Forces acting on the explicit liquid droplet

## 2.2. Evaporation Mechanism

As soon as the droplet exits into atmosphere, it starts evaporating. Let's assume that, the liquid droplet mainly consists of water and media inside is uniform, then suggesting to Fick's first law, vapor concentration flux $\vec{J}$ is given by:

$$\vec{J} = D\nabla \vec{c} \quad (5)$$

where, $D$ is a diffusion coefficient and $c$ is the water vapor concentration. Total flux should be mass change rate, which for the spherically symmetric case can be written as:

$$J \cdot 4\pi r^2 = -\frac{dm}{dt} \quad (6)$$

After combining the above equations and integrating both sides with appropriate limits (from the surface of the droplet to infinity - a room), gives us the mass change rate and further integration of the equation with respect to time leads to the well-known droplet mass-time dependence:

$$m_{(t)} = \frac{4}{3}\pi\rho_{in}\left(R_0^2 - \frac{D\Delta c}{\rho_{in}}t\right)^{3/2} \quad (7)$$

where $\Delta c$ and $c_R - c_\infty$ is the surface-room concentration difference. When air is fully dry $c_\infty = 0$, but as it gets more humid, $c_\infty$ also increases, reducing the concentration gradient, which in turn lowers the mass/radius change rate. It is worth mentioning that if $c_\infty > c_R$, then the droplet will start to enlarge and gain more mass instead of evaporating. Finally, with this on hand, we obtain a set of differential equations, governing the overall dynamics of the droplets which can be solved numerically for given initial conditions. In Figure 2, we study the trajectory characteristics obtained from the coupled differential equations solved for the following initial conditions: $R_0 = 200\,\mu m$ and $\upsilon_0 = 10\,m/s$. It is evident, that air induced drag has a significant role in the rate of velocity change and its decrease overtime, results in a reduced respiratory distance coverage.

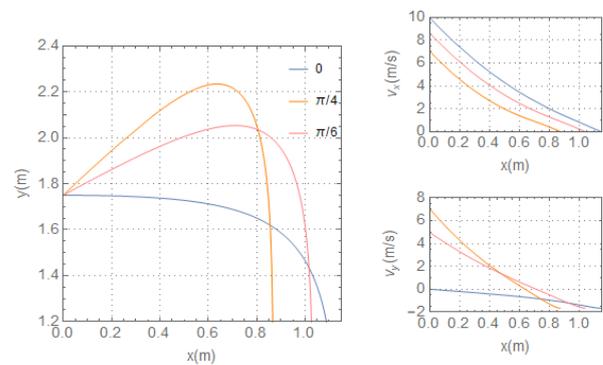

Figure 2: Trajectory and the axial velocity evolution. the set of parameters is: $R_0 = 200\,\mu m$ and $\upsilon_0 = 10\,m/s$ (initial conditions for variable hypothetical sample angles).

## III. MODELING

During the respiration, there is expelled the whole spectrum of droplets with different radii. Some might reach destination before complete evaporation or vice versa, completely evaporate midway or settle to ground due to gravity. To account for the all-case scenario, it is significant to know the probability of forming those particles. In [10], the author has studied the size distribution of respiratory droplets for sneezing, coughing with mouse open/closed and speaking loudly (Fig.3). By analyzing the data one can straightforwardly derive the size probability density function:

$$p = \frac{1}{R_0 \sigma \sqrt{2\pi}} \exp\left[-\frac{(\ln(R_0)-\psi)^2}{2\sigma^2}\right] \quad (8)$$

where, $R_0$ is the initial radius of the droplet, $\psi$ and $\sigma$ are the mean value and the standard deviation of the natural logarithm of radius respectively. Their corresponding numerical values are given on Table 1 for each kind of respiration. Apparently, the most frequent radii vary in the range $25 - 50 \mu m$.

Table 1: Log-normal distribution parameters

|   | Sneeze | Cough with mouth closed | Cough with mouth open | Speaking loudly |
|---|---|---|---|---|
| $\psi$ | 3.976 | 3.937 | 4.356 | 4.211 |
| $\sigma$ | 0.929 | 0.926 | 0.902 | 0.928 |

Likewise, the study of [10], in the present model we investigate the propagation of 3000 generated droplets from the orifice, by taking into account the above-mentioned distribution, at the initial vertical coordinate $y_0 = 1.75m$, with the initial velocity (10m/s) and a randomly distributed inclination angle of velocity towards the horizontal in the interval $[-\pi/6, \pi/6]$.

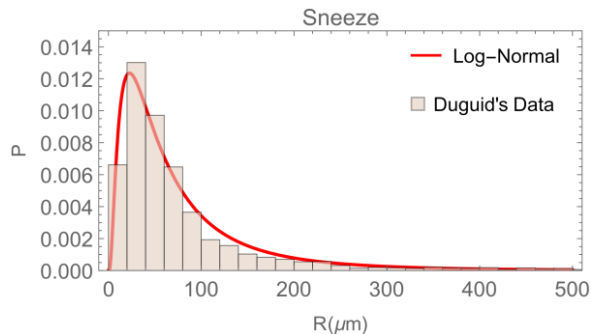

Figure 3: Duguid's data and respective log-normal distribution curve for sneezing and coughing. The most frequent radius appears to be from 25-50$\mu m$.

## IV. RESULTS AND DISCUSSION

In this part we further analyze and discuss the results compiled from previous sections. First of all, let's make a distinction between contagious and non-contagious particles: due to a randomly generated angle, it is given that some are bound to head straight towards the ground, which obviously do not pose any threat. As we study mouth-to mouth transmission, we consider the droplets to be dangerous if they reach a certain mean face level during the motion.

Evaporation of exhaled solution is bound to depend on initial batch of various components (e.g., salts, mucus), but should also be affected by ambient parameters such as for example temperature and humidity. In (7) mass change rate is directly influenced by water vapor concentration gradient, which would also vary for different values of the relative ambient humidity (RH) for $T_\infty = 295 K$.

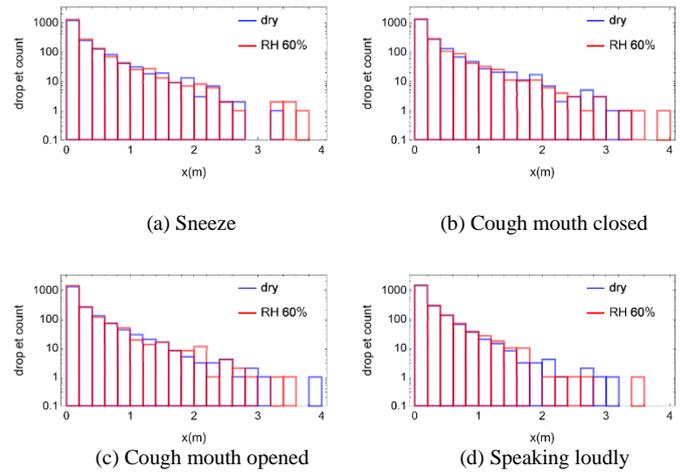

(a) Sneeze　　　　　　　(b) Cough mouth closed

(c) Cough mouth opened　　　(d) Speaking loudly

Figure 4: Histograms demonstrate the count of droplets over a horizontal distance at the mean-contagious height. Initial droplet velocities for each type of respiration are following: (a) sneezing 100 m/s, (b) coughing mouth closed 40 m/s, (c) coughing mouth open 15 m/s, (d) speaking loudly 5m/s.

Histograms in Figure 4 show a quantity of infectious droplets which manage to reach a certain distance. For a cough (with initial particle velocity $\upsilon_0 = 10 m/s$), compared to a completely dry air, 2.5% and 6.6% more particles got grounded under 0.5$m$ checkpoint at 30% and 60% RH respectively. Further the distance-less the number of particles in infectious area, but histograms in Fig.4 still do not describe the danger properly, since it only shows the certain count of droplets, which does not indicate the exact amount of viral load. To briefly account for that we add up the volume of the droplets for each histogram bar to obtain the distance-total volume dependence. That would be more accurate to describe the contagiousness risk, since the pathogen number will be proportional to the total volume. According to [11], a typical infected coughing individual-" high emitter" produces 36030 copies per $cm^3$. If we combine that with the simulated viral load, we can approximate number of viral copies, which varies from hundreds to several thousands.

If we take a closer look at Fig.4, one would notice a significant remnant of the viral load after 2m marker. Mainly, larger droplets contain more viral load making us prone to the powerful respiration even beyond that of the given recommendation by experts.

Those particles almost instantaneously reach a face area, which is extra dangerous for a person directly exposed to a coughing individual. An airborne time of such droplets seem to vary for the different values of ambient relative humidity. For higher RH evaporation process slows down, which also affects surface area change resulting in extra drag force and less airborne time (Fig.5). Despite the evaporation, a risk of infection might be increased, since the remnants of dry viral load are being exposed to atmosphere, therefore the process will be enhanced by outer air currents and a transportation mechanism will be changed, which is yet a completely different problem to investigate.

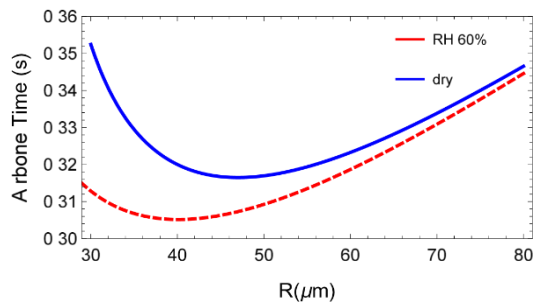

Figure 5: ($\alpha = \pi/6$, $\upsilon_0 = 10\,m/s$) airborne time of test-sample droplet for different ambient RH

## V. CONCLUSION AND FUTURE SCOPE

We attempted to study characteristics of dynamics of liquid droplets, which has given us a different perspective about the transportation of infectious particles. Detailed analysis of the evaporating droplet's momentum gives us a general idea about propagation process. As evident from Figure 2, velocity proportional drag force plays quite a role in such events, due to which most of large-size particles ($R_0 > 30\mu m$) quickly enter the motion mode with not significant horizontal velocity component and fall below the contagious area within the regulation range. We also observed slight changes for different ambient relative humidity resulting in less airborne time (Fig.5) and significant change in viral load (Fig.4), but rapid evaporation will have even larger contribution in transporting spray-like respiratory fluid (droplets with $R_0 < 10\mu m$), where motion becomes significantly complex due to the existent turbulent flow currents.

Direct inhalation of such particles poses a great threat to our health, since they contain the viral load in such large quantities. We would like to encourage the following social distancing standards, but there is also need of caution, since it is evident that $2m$ checkmark is not always a safe distance of total invincibility, with all that transmitted significant amount of viral load, to cause a possible infection. But in this part, the problem has to be considered from the biology point of view to make a final verdict.

## ACKNOWLEDGMENT

The authors are grateful to prof. I. Pipia for valuable comments. The research of ZU was supported by the Knowledge Foundation at Free University of Tbilisi.